\documentclass{aa}  
\usepackage{natbib}
\usepackage{graphicx}
\usepackage{txfonts}
\usepackage[normalem]{ulem}
\usepackage{xcolor}
\usepackage{placeins} 

\usepackage{caption}
\usepackage{placeins} 

\definecolor{links}{rgb}{0.11, 0.67, 0.84}
\usepackage[colorlinks=true,allcolors=links]{hyperref}

\newcommand{\rhodtfe}{$\rho_\mathrm{DTFE}$}
\newcommand{\LOD}{$1 + \delta_\mathrm{DTFE}$}
\newcommand{\Nstream}{$N_\mathrm{streams}$}

\begin{document} 

  \title{Flows around galaxies. I.  The dependence of galaxy connectivity on cosmic environments and effects on the star formation rate}
  
  \titlerunning{Galaxy connectivity, cosmic environments, and star formation rate}
  
  \author{Daniela Gal\'arraga-Espinosa\inst{} 
  \and Enrico Garaldi\inst{}
  \and Guinevere Kauffmann\inst{}  
        }

  \institute{Max-Planck Institute for Astrophysics, Karl-Schwarzschild-Str.~1, D-85741 Garching, Germany\\
     \email{danigaes@mpa-garching.mpg.de}}

  \date{Received XXX; accepted YYY}

  \abstract{
   {With the aim of bringing substantial insight to the fundamental question of how galaxies acquire their material for star formation, we present the first comprehensive characterisation of the galaxy connectivity (i.e. the number of small-scale filamentary streams connected to a galaxy) in relation to the cosmic environment, and a statistical exploration of the impact of connectivity on the star formation rate (SFR) at $z=2$. We detected kiloparsec-scale filaments directly connected to galaxies by applying the DisPerSE filament finder to the dark matter density around 2942 central galaxies ($M_* > 10^{8}$ $\mathrm{M}_\odot / h$) of the TNG50-1 simulation. Our results demonstrate that galaxy connectivity spans a broad range (from 0 to 9), with more than half of the galaxies connected to two or three streams. We examined a variety of factors that might influence the connectivity and found that it increases with mass, decreases with local density for low-mass galaxies, and does not depend on local environment, estimated by the Delaunay tessellation, for high-mass galaxies. 
   Beyond mass and local density, we further classified galaxies according to their location in different cosmic web environments, and we highlight the influence of the large-scale structure on the number of connected streams. Our results reflect the different strengths of the cosmic tides, which can prevent the formation of coherent streams feeding the galaxies or even disconnect the galaxy from its local web. 
   Finally, we show that at fixed local density, the SFR of low-mass galaxies is up to $5.9\sigma$ higher as a result of connectivity. This SFR boost is even higher ($6.3\sigma$) for galaxies that are embedded in cosmic filaments, where the available matter reservoirs are large. A milder impact is found for high-mass galaxies, which indicates different relative efficiencies of matter inflow via small-scale streams in galaxies of different masses.}}

\keywords{(cosmology:) large-scale structure of Universe, galaxies: evolution , galaxies: star formation, galaxies: statistics, methods: numerical, methods: statistical}

\maketitle
\newpage


\section{Introduction}

Under the action of gravity, matter on large scales in the Universe is assembled to form a gigantic network composed of nodes, filaments, walls, and voids. This is called the cosmic web \citep{Lapparent1986, Bond1996}. Emerging from the initial density fluctuations \citep{Zeldovich1970}, this cosmic skeleton is mainly composed of and ruled by the dynamics of dark matter (DM). Driven by gravity, baryonic matter falls into the DM potential wells.
The structure of the cosmic web is highly multi-scale \citep{AragonCalvo2010}. While the nodes of the web, hosting the most massive galaxy clusters, are connected to large-scale cosmic filaments with widths of several megaparsec \citep[e.g.][]{Gouin2021, Gouin2022, GalarragaEspinosa2022}, small haloes are also attached to the web via smaller-scale filaments that are characterised by widths of tens of kiloparsec \citep[e.g.][]{Ramsoy2021}. These small-scale filaments, or streams, are expected to have a strong effect on the evolution and properties of galaxies residing at the centre of these haloes. 

Galaxies are thought to be formed at the intersection of these small-scale filamentary streams, which, in theory, feed the galaxies with the cold and dense material necessary for star formation \citep[e.g.][]{Birnboim2003_streams, Keres2005, Ocvirk2008, Dekel2009_coldstreams, Pichon2011_streams_and_discs, FaucherGiguere2011_letter, FaucherGiguere2011_streams_and_outflows, Danovich2012_streams}. The theoretical prediction is thus that these filaments act as highways of matter, from the large-scale reservoirs down to the halo centres.
This picture is supported by studies in observations such as \cite{Bauermeister2010_GasConsumption}, and more recently, \cite{Prescott2015_Lya_largescales} and \cite{Zabl2019_Lya}, who have clearly demonstrated the need of gas replenishment from external reservoirs. Nevertheless, 
other processes can also participate in the fuelling of the galaxy with the material for star formation. These are, for example, the precipitation of hot gas in virial equilibrium with the dark matter halo \citep{Keres2005}, the recycling of gas from the circum-galactic medium (CGM), or even galaxy-galaxy mergers, which drive gas from the outskirts of galaxies into their centres, where they form stars very rapidly in a so-called starburst. 
While \cite{Stewart2017} has proven that gas accretion into haloes via filamentary streams is a robust prediction of $\Lambda$-CDM (because it is independent of the adopted code and feedback model), \citet{Nelson2013_movingmesh} has shown that gas transport inside haloes, that is, from the CGM into the galaxies, is strongly impacted by the numerical scheme of the hydrodynamical simulation (which 
alters the relative importance of accretion via cold streams and via cooling of shock-heated gas).
Thus, the question of how galaxies acquire the material for star formation and the relative efficiency of the processes involved is yet to be understood. 

Another active topic of investigation is why galaxies stop forming stars. The current picture involves a complex variety of feedback and environmental processes that regulate the balance between gas inflows and outflows around  galaxies, and whose relative impact strongly depends on other parameters such as galaxy mass and environment \citep[e.g.][]{Kauffmann2004, Baldry2004, Bamford2009_GalaxyZOO, Peng2010_QuenchingGalaxies, Moutard2018_quenching}.
star formation could be suppressed either by internal mechanisms such as energetic feedback from supernovae or accreting black holes, or by environmental effects such as ram-pressure stripping or tidal interactions. The latter are external processes, which according to \cite{AragonCalvo2019_CWdisconnection}, are fundamentally linked with the disconnection (or detachment) of the galaxy from its filamentary streams. This engenders a mechanical starvation either by removing gas reservoirs or by preventing gas from reaching galaxies. 

In this context, it is crucial to re-evaluate the relative effect of filamentary streams on galaxy evolution in a cosmological context, that is, to take the environment in which galaxies form and evolve into account. A study in a cosmological context is crucial because it is now well established, both in observations and simulations, that beyond the trends with mass and local environment, galaxy properties also vary as a function of their location in the structures of the cosmic web. For example, galaxies located in cluster environments are more massive, form fewer stars, are redder, and their morphologies are more elliptical than those in less dense regions \citep[see e.g. the reviews of][]{Dressler1980, BoselliGavazzi2006, BoselliGavazzi2014}. Similar trends are found in the cores of cosmic filaments with respect to regions that lie farther away from the spines (e.g. \citet{Pandey2006, Alpaslan2016_GAMAgalaxies, Chen2017_fil_gal, Malavasi2017, Laigle2018, Bonjean2020filaments, Rost2020, Welker2020_sami, Gouin2020, Winkel2021_galaxies} in observations and \citet{Kraljic2018, Kraljic2019, GaneshaiahVeena2018_CosmicBalletI, GaneshaiahVeena2021_CosmicBalletIII, Malavasi2022_spin} in simulations).\\

This paper is the first in a series providing an updated picture of the impact of filamentary flows on galaxy evolution. We use the recent TNG50 simulation \citep{Pillepich2019_TNG50, Nelson2019_TNG50} to perform a statistical analysis of the number of (kiloparsec-scale) streams connected to galaxies, hereafter referred to as the galaxy connectivity, as a function of the environment of the galaxy in the cosmic web (defined at megaparsec scales). While potential inflows and outflows of baryons along these streams will be studied in the second part of this project, we provide in this paper a first exploration of the impact of galaxy connectivity on the specific star formation rate (sSFR), defined as the SFR normalised by galaxy stellar mass.
We emphasise that the multi-scale analysis performed in this work is different from previous studies, which have rather focused on how large-scale structures, such as groups or clusters, are connected to large-scale cosmic filaments on megaparsec scales \citep{Kraljic2020, Gouin2021, Gouin2022}, yielding relevant conclusions on the properties of the cosmic environments where galaxies live, but not on the properties of galaxies themselves. Moreover, we note that this type of study has only recently been enabled through the advent of large-scale hydrodynamical simulations with more robust baryonic models and increasing resolution \citep[e.g.][]{Tremmel2017_Romulus, Pillepich2019_TNG50, Dubois2021_NewHorizon}, and is crucial in order to interpret future observations.

This paper is organised as follows. Section~\ref{Sect:Data} introduces the TNG50 simulation and the dataset of galaxies. We present the detection of the small-scale filamentary streams as well as the large-scale cosmic web in Sect.~\ref{Sect:MethodFils}. Results about galaxy connectivity are first introduced in Sect.~\ref{Sect:GalConnectivity}, and the impact of the large-scale environments is discussed in Sect.~\ref{Sect:envs_LSS}. Finally, the relation between connectivity and SFR is explored in Sect.~\ref{Sect:SFR}, and we summarise our conclusions in Sect.~\ref{Sect:Conclusions}. Throughout this paper, we adopt the values of the cosmological parameters given by \cite{Planck2015Cosmo}, that is,  $\Omega_{\Lambda,0} = 0.6911$, $\Omega_{m,0}=0.3089$, $\Omega_{b,0}=0.0486$, $\sigma_{8}=0.8159$, $n_s=0.9667,$ and $h=0.6774$. The error bars correspond to the errors on the mean values, derived from bootstrap resampling.

\section{\label{Sect:Data}Data}

\subsection{\label{SubSect:IllustrisTNG}TNG50-1 simulation}

The analysis presented in this work uses the outputs of the TNG50-1 simulation, which is the box of the gravito-magnetohydrodynamical simulation suite, IllustrisTNG\footnote{\url{https://www.tng-project.org}} , with the highest resolution \citep{Pillepich2018TNGmodel, Nelson2019_TNGdata_release, Pillepich2019_TNG50}. 
With a mass resolution of $m_\mathrm{DM} = 3.07 \times 10^5 \, \mathrm{M}_\odot /h$ and a volume of $(35 \, \mathrm{cMpc/h})^3$, this box is adapted to study the small-scale (kiloparsec) filamentary streams in a statistical way. We note that the IllustrisTNG project was run with the moving-mesh code Arepo \citep{Arepo}, and the baryonic models and prescriptions were specifically calibrated on observational data to match the observed galaxy properties and statistics \citep{Pillepich2018TNGmodel, Nelson2019_TNGdata_release}.
All the following results are derived from the TNG50-1 snapshot at redshift $z=2$. This redshift typically corresponds to the so-called cosmic noon, the epoch in which galaxies formed stars most actively, which is therefore the ideal time at which to examine galaxy connectivity and its influence on star formation.

In the future, we will build on the current work by investigating the gas content of the DM filaments identified here. Therefore, it is crucial to verify that the simulation is also suited for this objective. We verified that TNG50-1 meets the resolution criterion found by \cite{Ramsoy2021} for capturing the filament physical properties (e.g. the shocks in their temperature profile).

\subsection{\label{SubSect:GalaxySelection}Galaxy selection}

From the subhalo catalogue of the TNG50-1 simulation at $z=2$ (produced using the Subfind code \citet{Springel2005}), we selected the central objects with stellar masses higher than $M_* > 10^{8}$ $\mathrm{M}_\odot / h$. The maximum subhalo stellar mass is $4 \times 10^{11} \, \mathrm{M}_\odot / h$. This selection in mass chooses subhaloes at $z=2$ that will most likely become systems with a typical mass of $10^{9}-10^{12}$ $\mathrm{M}_\odot / h$ at $z=0$ \citep{Brinchmann2004, Taylor2011}.

Importantly, we emphasise that we focus on central galaxies alone. They are identified as the subhaloes at the centre of their corresponding friends-of-fiends (FoF) halo. Satellite galaxies were excluded from this analysis because we found (visually) that they lie very close to the spine of the filaments associated with their central galaxy, that is, satellites are probably part of these streams. A more quantitative analysis of satellite galaxies and their position relative to the filamentary streams will be performed in a future work.

In addition, in order to facilitate the procedure of extracting the filamentary streams (see next section), we conservatively chose to discard the central galaxies located at distances smaller than 1.5 cMpc/h from the edges of the full simulation box. 
We finally note that $98.8 \%$ of the remaining galaxies in our catalogue are star forming, as shown by their main sequence in the $M_* - \mathrm{SFR}$ plane presented in Appendix~\ref{Appendix:mainsequence}. We discarded the few passive galaxies (35) so that the analysis presented in this work does not mix two different galaxy populations (i.e. galaxies at different evolutionary stages) at $z=2$. Based on the selections presented above, the total number of galaxies analysed in this work is 2942.

\section{\label{Sect:MethodFils}Finding small- and large-scale filaments}

In this section, we explain the procedure we adopted to extract the small-scale streams connected to galaxies and the large-scale (megaparsec) cosmic web skeleton. To detect these multi-scale structures with an optimal resolution, we employed the filament finder DisPerSE (Sect.~\ref{SubSect:Disperse}) to adapted regions of the DM density field. The small-scale streams were detected from selected sub-boxes centred on the position of individual galaxies (see Sect.~\ref{SubSect:Extracting_streams}), and the entire simulation box was used to find the large-scale filamentary skeleton, as explained in Sect.~\ref{SubSect:Extracting_CW}.

\subsection{\label{SubSect:Disperse}Filament extractor code DisPerSE}

DisPerSE \citep{Disperse_paper1, Sousbie2011b} is a publicly available code that detects the cosmic skeleton from the topology of the density field (e.g. the DM density), using the discrete Morse theory and the theory of persistence \citep[see][and references therein]{Disperse_paper1}. This algorithm identifies the critical points of the field, that is, the points with a vanishing density gradient. Filaments are defined as the ridges of the density field connecting maximum-density critical points (hereafter CPmax) to saddles\footnote{For an illustration, we refer the reader to Fig.~2 of \cite{GalarragaEspinosa2020}}. Importantly, the minimum significance of the detected filaments with respect to the noise can be set by fixing the persistence threshold of the corresponding pairs of CPmax-saddle critical points. For density fields that are computed on regular grids (e.g. in this work), the persistence threshold needs to be set via the \textit{cut} parameter. The value of this parameter should correspond to the amplitude of the noise of the input density grid, so that any CPmax-saddle critical pair with density difference lower than the adopted threshold is rejected.
For further details, we refer to the DisPerSE presentation papers \citep{Disperse_paper1, Sousbie2011b} and website\footnote{\url{http://www2.iap.fr/users/sousbie/web/html/indexd41d.html}}.

\subsection{\label{SubSect:Extracting_streams}Extracting the small-scale streams}

\begin{figure*}
    \centering
    \includegraphics[width=0.7\textwidth]{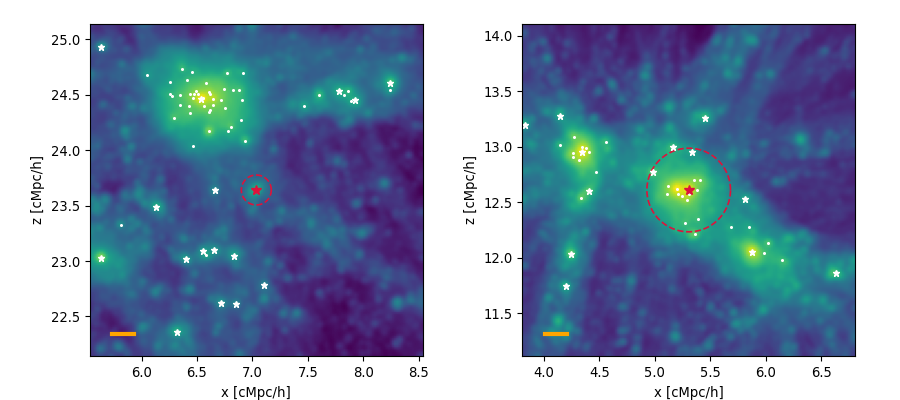}
    \includegraphics[width=0.7\textwidth]{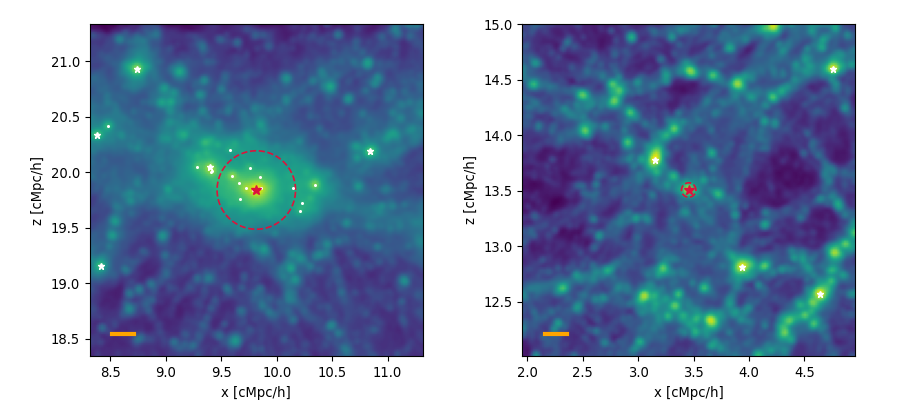}
    \includegraphics[width=0.7\textwidth]{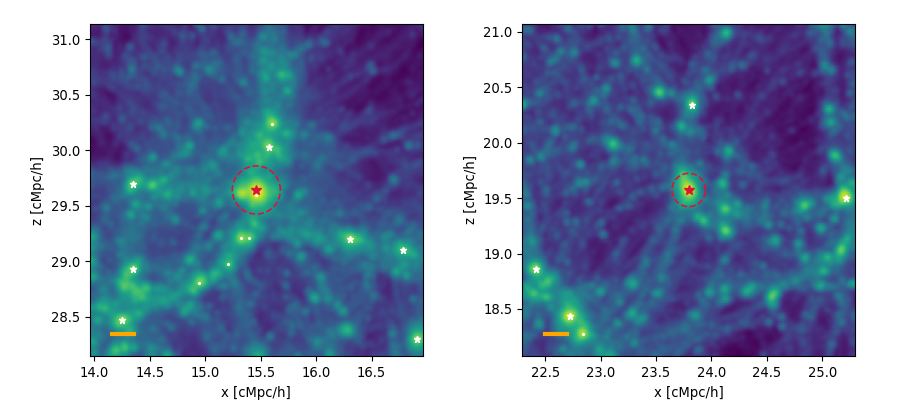}
    \caption{Examples of 2D projected DM density fields. For each sub-box with a side of 3 cMpc/h, the red star and red circle correspond to the analysed central galaxy of mass $M_* > 10^{8} \, \mathrm{M}_\odot$ and to the $R_{200}$ radius of its host FoF halo, respectively. The small white stars and white dots indicate other centrals and satellite galaxies located in the sub-box, respectively. The length of the orange line in the bottom left part of the panels corresponds to ten times the resolution scale of the grid chosen to project the DM density and extract the skeleton, i.e. ten times 20 ckpc/h.} 
    \label{Fig:2d_DENSITIES}
\end{figure*}

\begin{figure*}
    \centering
    \includegraphics[width=0.8\textwidth]{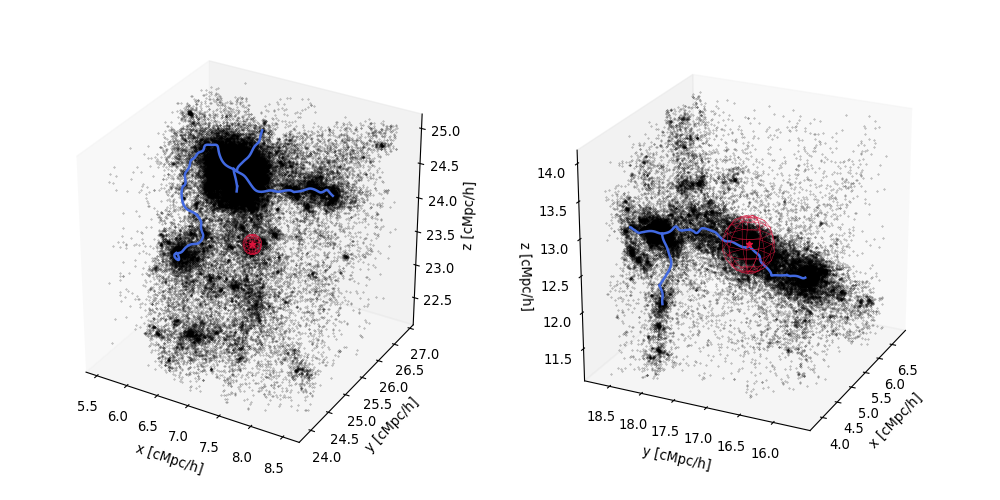}
    \includegraphics[width=0.8\textwidth]{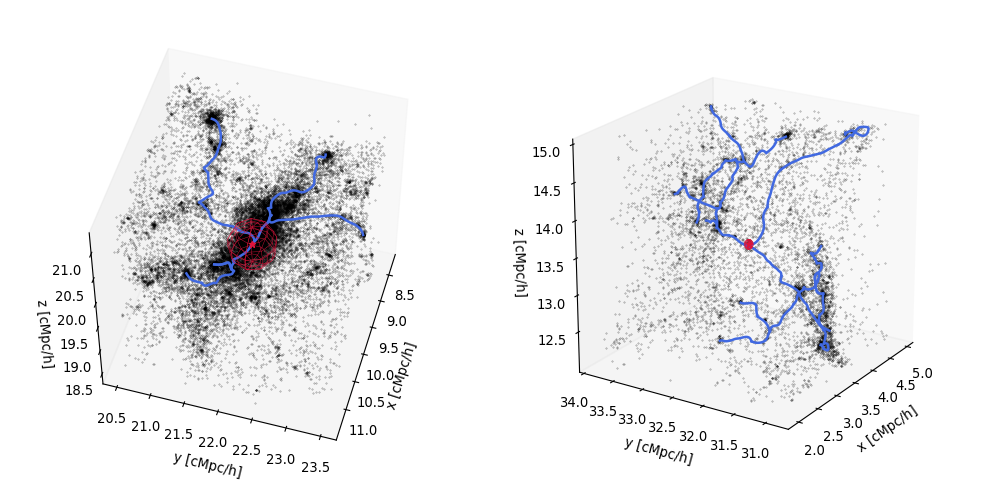}
    \includegraphics[width=0.8\textwidth]{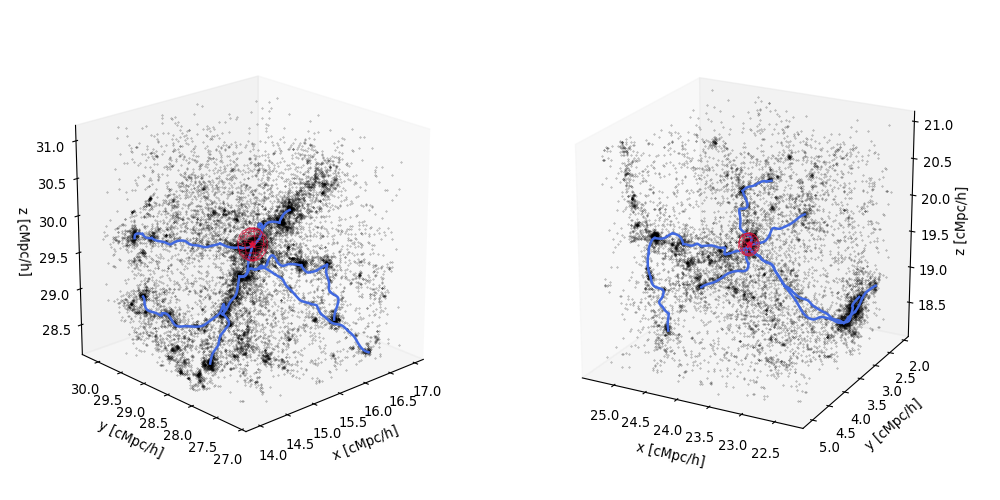}
    \caption{Examples of the resulting small-scale filamentary streams detected with DisPerSE (blue lines) for the same galaxies as in Fig.~\ref{Fig:2d_DENSITIES}. The red spheres correspond to spheres with a radius $R_{200}$ of the galaxy host FoF halo. For illustration, the black points represent a random sub-sample (1/1000) of the DM particle distribution in the sub-box, but the filaments were detected from 3D grids of the DM density field, as described in Sect.~\ref{SubSect:Extracting_streams}.}
    \label{Fig:3d_STREAMS}
\end{figure*}

We detected the small-scale (kiloparsec) filamentary streams connected to galaxies by applying DisPerSE to the local DM density field.
For each individual galaxy, we selected the DM particles in sub-boxes with a side $L = 3$ cMpc/h centred on the position of the galaxy. This value was chosen in order to capture the galaxy environment beyond the typical scales of the CGM, thus probing the large-scale matter distribution. For reference, 3 cMpc/h is a factor of five larger than the largest virial radius of the haloes of the galaxies in our catalogue. We also verified that increasing the size of the sub-boxes did not change the galaxy connectivity estimates. This analysis is presented in Appendix~\ref{Appendix:LargerBox}.

The DM density field was computed by projecting the particles inside the galaxy sub-box onto a regular grid of $N_\mathrm{pix}=150$ pixels per side. We applied a Gaussian filter with a standard deviation equal to the size of a pixel (i.e. $3/150 = 20$ ckpc/h) to the grid values, and we rescaled the resulting pixel values by the standard deviation. These steps enable the application of DisPerSE with the same parametrisation to all the 2942 density grids. Figure~\ref{Fig:2d_DENSITIES} presents some examples of DM density grids (projected along the \textit{y}-axis). For each panel, the analysed galaxy (red star) is at the centre of the sub-box, and the virial radius of the host halo is indicated by the red circle. Other centrals and satellites located in the same sub-box are shown as white stars and dots, respectively.

DisPerSE was then applied to each one of the 2942 regular grids, so that each galaxy possessed its own set of small-scale filaments.
We treated the non-periodic boundary conditions of each sub-box by specifying the \textit{periodicity 0} keyword in the computation of the Morse-smale complexes. The persistence threshold, which acts as a filter of the features that are likely to have been generated by noise, was determined by exploring a broad range of values of the DisPerSE \textit{cut} parameter. 
In Appendix~\ref{Appendix:Disperse_cut} we assess the impact of this parameter on the final number of streams connected to the galaxies. We show that \textit{cut} values above 25 are required to efficiently remove filaments arising from the noise, and that the progressive increase in persistence beyond \textit{cut}=25 only mildly impacts the connectivity estimates (by slightly lowering the connectivity normalisation). From this study, we conclude that provided the noise-induced filaments are removed, our statistical results on connectivity depend only very weakly on the exact value of the DisPerSE persistence threshold. We therefore chose to fix this parameter to \textit{cut}=30 after visual inspection of several random galaxies. This threshold kept some small filament portions that visually agreed well with the underlying DM density field, but were absent in the skeletons derived with higher \textit{cut} values. 

The positions of the resulting streams were then smoothed using the DisPerSE \texttt{skelconv} function. By straightening the skeleton segments and smoothing sharp and possibly non-physical edges between them, this final step alleviates the effect of shot noise on the geometry of the filaments. This procedure does not affect the topology of the density field \citep{Codis2018}, and thus keeps the connectivity unchanged. Figure~\ref{Fig:3d_STREAMS} presents some examples of the resulting streams in 3D boxes. These correspond to the same galaxies as in Fig.~\ref{Fig:2d_DENSITIES}.\\

It is worth noting that we explicitly chose not to identify the filaments from the DM particle distribution in order to avoid the inevitable contamination from small clumps (at scales $< 10$ kpc) and, most of all, from the high shot-noise levels provoked by the great number of DM particles. We found that skeletons detected in the particle distribution were extremely sensitive to the slightest changes of persistence threshold, causing filaments to appear and disappear, and provoking radical changes in the position of even the most prominent structures.
It is therefore a more stable method to run DisPerSE on a DM density grid, but this has the drawback of setting an intrinsic resolution scale, $L/N_\mathrm{pix} = 20$ ckpc/h (see the orange lines in Fig.~\ref{Fig:2d_DENSITIES}, which correspond to ten pixels). This means that the positions of the filament spines are determined with a precision of $\pm 10$ ckpc/h. While this precision limit might compromise the accuracy of radial density profiles (because the exact position of the filament cores is uncertain), we emphasise that it does not undermine the results on connectivity we present here.

\subsection{\label{SubSect:Extracting_CW}Extracting the large-scale cosmic skeleton}

\begin{figure}
    \centering
    \includegraphics[width=0.47\textwidth]{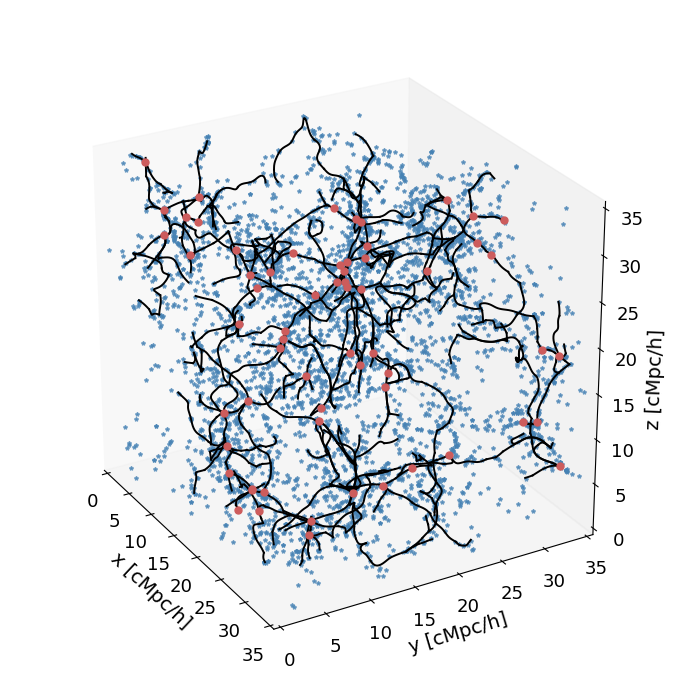}
    \caption{Cosmic filaments (black lines) of the full TNG50-1 simulation. The red dots show the position of clusters of galaxies, i.e. the FoF haloes with masses $M_{200}>10^{12} \, \mathrm{M}_\odot/h$. The blue stars correspond to the 2942 central galaxies analysed in this work. The detection of the cosmic skeleton is presented in Sect.~\ref{SubSect:Extracting_CW}.}
    \label{Fig:3d_LSS}
\end{figure}

With a similar method as for the small-scale streams, the large-scale (megaparsec) cosmic skeleton was detected by projecting the full TNG50-1 DM particle distribution onto a regular grid of 150 pixels per side, yielding an intrinsic resolution scale of $35/150 = 0.23$ cMpc/h for these large-scale cosmic filaments. 
The persistence threshold was set after analysing the outputs obtained with different values of the \textit{cut} parameter. For the large-scale structure, a physical criterion for determining the robustness of the skeleton is that the DisPerSE CPmax points match the positions of the most massive haloes, such as those of groups and clusters of galaxies. The results of this matching is presented in Appendix~\ref{Appendix:Disperse_cutLSS}, in which the choice running of DisPerSE with a persistence threshold of 6 is also justified. Figure~\ref{Fig:3d_LSS} shows the resulting cosmic filaments in the 3D box of the TNG50-1 simulation.
We recall that the identification of the large-scale cosmic skeleton in this work is done solely with the aim of classifying the galaxies into different cosmic environments, as we show in Sect.~\ref{Sect:envs_LSS}.

\section{\label{Sect:GalConnectivity}Galaxy connectivity}

After detecting the small-scale filamentary streams, we present a statistical analysis of the galaxy connectivity in this section, that is, the number of streams to which each galaxy is connected. Sect.~\ref{SubSect:Nstream_general} presents general results for all the galaxies, and secondary dependences on galaxy mass and local environment are analysed in Sect.~\ref{SubSect:TrendsMass} and Sect.~\ref{SubSect:LOD}, respectively.

\subsection{\label{SubSect:Nstream_general}General results}

\begin{figure}
    \centering
    \includegraphics[width=0.5\textwidth]{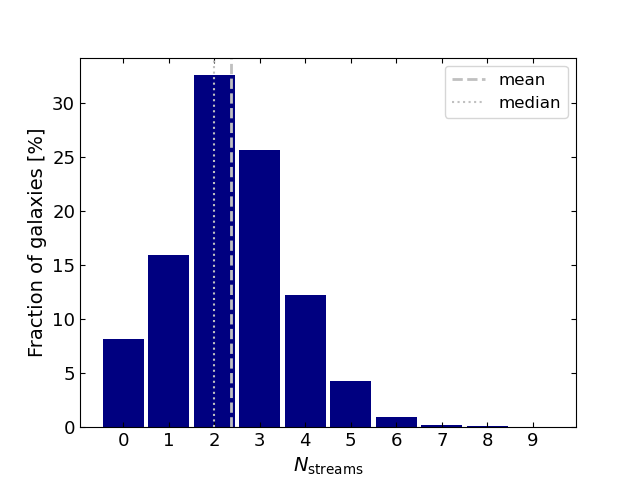}
    \caption{Histogram of the number of streams, \Nstream, connected to each galaxy of the full galaxy catalogue. The vertical dotted and dashed lines show the mean and median values, respectively. The number of streams to which a galaxy is connected defines the galaxy connectivity.}
    \label{Fig:Nstream0}
\end{figure}

Figure~\ref{Fig:Nstream0} shows the distribution of the number of streams to which a galaxy is connected for all the 2942 galaxies of the dataset. This number was obtained by counting the number of filaments that cross the virial radius of the host haloes. This figure shows that the galaxy connectivity spans a broad range (from 0 to 9) and presents a long tail towards high connectivity values, indicating that high connectivity is possible, but occurs quite rarely. The highest peaks are seen for $N_\mathrm{streams} =2$ and 3, with $32.5\%$ and $25.6 \%$ galaxies connected to two and three streams, respectively. The mean and median values of the distribution are 2.36 and 2, respectively.

The skewed shape and broad range of the distribution presented in Fig.~\ref{Fig:Nstream0} indicate that additional factors may affect the galaxy connectivity. In the next sections, we therefore distinguish secondary dependences on galaxy mass and local environment.

\subsection{\label{SubSect:TrendsMass}Trends with galaxy mass}

\begin{figure}
    \centering
    \includegraphics[width=0.5\textwidth]{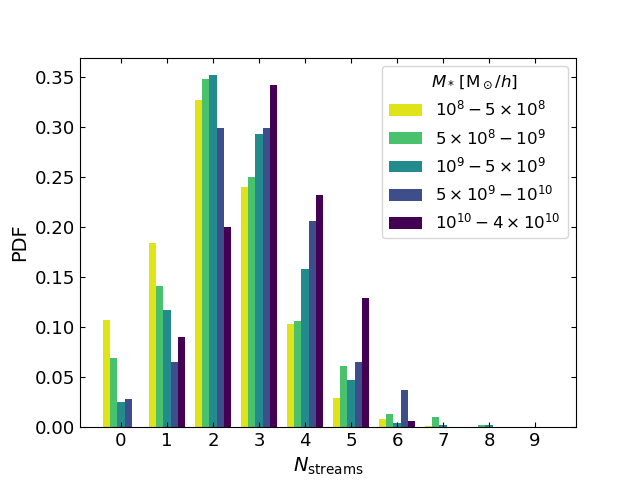}
    \caption{Connectivity distribution by bins of galaxy mass (from yellow to purple in increasing mass). The mean value of each distribution is $2.18 \pm 0.03, 2.49 \pm 0.07, 2.63 \pm 0.06, 2.93 \pm 0.12,$ and $3.13 \pm 0.10$ from the lowest to the highest masses.}
    \label{Fig:Nstr_dist_5binsM}
\end{figure}

In this section, we investigate the effects of the galaxy mass on its connectivity. Figure~\ref{Fig:Nstr_dist_5binsM} shows the \Nstream~distribution for galaxies separated into five different bins of stellar mass. A clear trend emerges: the distributions of the highest-mass bins (e.g. purple) are shifted towards higher connectivity values than those of the lowest-mass bins (e.g. yellow). More massive galaxies are therefore more connected than lower-mass galaxies. This result provides an extension to lower masses of the trend that is well established in galaxy clusters on megaparsec scales \citep[][]{AragonCalvo2010, Codis2018, DarraghFord2019, Sarron2019, Malavasi2020_sdss, Kraljic2020, Gouin2021}.
The mean and median values of the distributions of Fig.~\ref{Fig:Nstr_dist_5binsM} also reflect the described trend.  From the lowest to the highest masses, the mean connectivity values are 
$2.18 \pm 0.03, 2.49 \pm 0.07, 2.63 \pm 0.06, 2.93 \pm 0.12,$ and $3.13 \pm 0.10$. According to \cite{Codis2018}, a higher connectivity is predicted for high-density peaks (massive galaxies in our context) because all the eigenvalues of the Hessian matrix (i.e. the matrix of the second derivatives of the density field) are equal in the vicinity of these peaks, thus describing a situation of local isotropy where all incoming directions become possible \citep[see also][]{Pichon_Bernardeau1999}.

We note that the separation into different mass bins allows us to better understand the asymmetric shape of the total \Nstream~distribution presented in Fig.~\ref{Fig:Nstream0}. The peak at \Nstream$=0$ is mostly clearly associated with the lowest-mass galaxies, whose distributions in Fig.~\ref{Fig:Nstr_dist_5binsM} are more skewed than those of the highest-mass bins.\\

\begin{figure}
    \centering
    \includegraphics[width=0.5\textwidth]{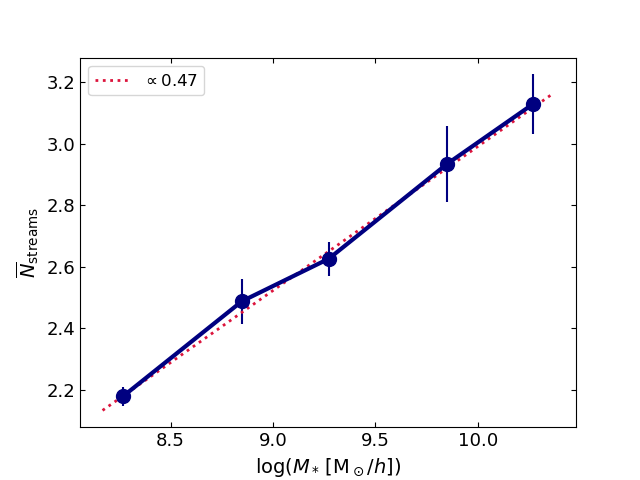}
    \caption{Relation between galaxy mass and mean connectivity, $\overline{N}_\mathrm{streams}$ (blue curve). The dashed red line shows the resulting fit-curve as presented in Eq.~\ref{Eq:fit}.}
    \label{Fig:meanNstr_binsM}
\end{figure}

To proceed in the quantitative analysis, we present in Fig.~\ref{Fig:meanNstr_binsM} the relation between the mean connectivity, $\overline{N}_\mathrm{streams}$, and galaxy mass. A simple logarithmic model was used to fit this relation, and the best-fit result is shown by the dashed red diagonal.The the resulting parameters are given by
\begin{equation}\label{Eq:fit}
    \overline{N}_\mathrm{streams} = (0.47 \pm 0.02) \times \log(M_* \,\, [\mathrm{M}_\odot/h]) - (1.69 \pm 0.13).
\end{equation}

We verified that the $\sim 0.5$ slope is independent of the number and limits of the mass bins. These results show that the trends of galaxy connectivity with mass can be captured quite well by a simple relation in the $\overline{N}_\mathrm{streams} - \log(M_*)$ plane. This relation echoes the theoretical results of \cite{Codis2018}, using peak theory.\\

In this section, we have shown that the number of streams connected to a galaxy depends on galaxy mass. We found the clear trend that more massive galaxies are more strongly connected than less massive galaxies on average. We now explore any dependences on the local environment of the galaxy, which is quantified by the local density.

\subsection{\label{SubSect:LOD}Trends with local density}

\begin{figure}
    \centering
    \includegraphics[width=0.5\textwidth]{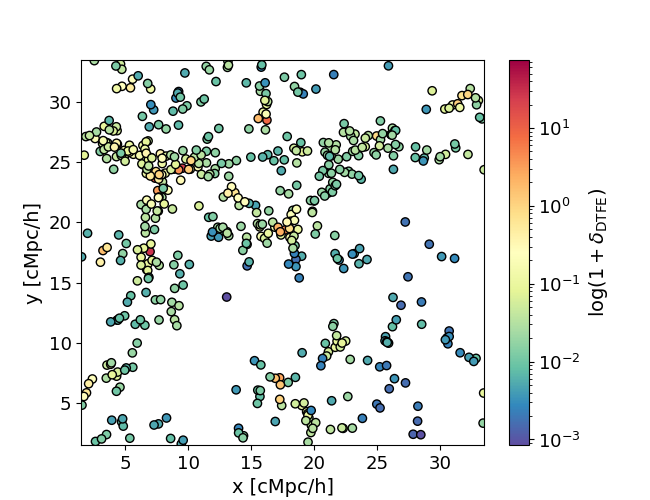}
    \caption{Example of  local over-density estimates for the galaxies (coloured circles) in a 5 cMpc/h thick slice of the TNG50-1 simulation.}
    \label{Fig:slice_OV}
\end{figure}

We used the Delaunay tessellation field estimator \citep[DTFE;][]{SchaapWeygaert2000, WeygaertSchaap2009} to compute the local densities of the galaxies. The DTFE was applied to the distribution of the 2942 massive centrals of our catalogue, so that each galaxy defined a vertex in the Delaunay tessellation and was attributed with a density value, hereafter \rhodtfe. In order to mitigate the effect of Poisson noise in our estimates, we smoothed the densities by averaging the value at each vertex with that of its direct neighbours in the Delaunay tessellation. After this smoothing, local over-densities were computed as
\begin{equation}
    1 + \delta_\mathrm{DTFE} = \frac{ \rho_\mathrm{DTFE} }{\langle \rho_\mathrm{DTFE} \rangle },
\end{equation}
where $\langle \rho_\mathrm{DTFE} \rangle$ represents the average of all the densities.
Physically, this quantity can be interpreted as a proxy for the crowding of the local environment of the galaxy. Galaxies in crowded regions (i.e. with many other neighbouring galaxies) are associated with high local over-densities, whereas low local over-densities pertain to galaxies living in more locally empty, less crowded spaces. This is clearly illustrated in the example of Fig.~\ref{Fig:slice_OV}.\\

\begin{figure*}
    \centering
    \includegraphics[width=0.5\textwidth]{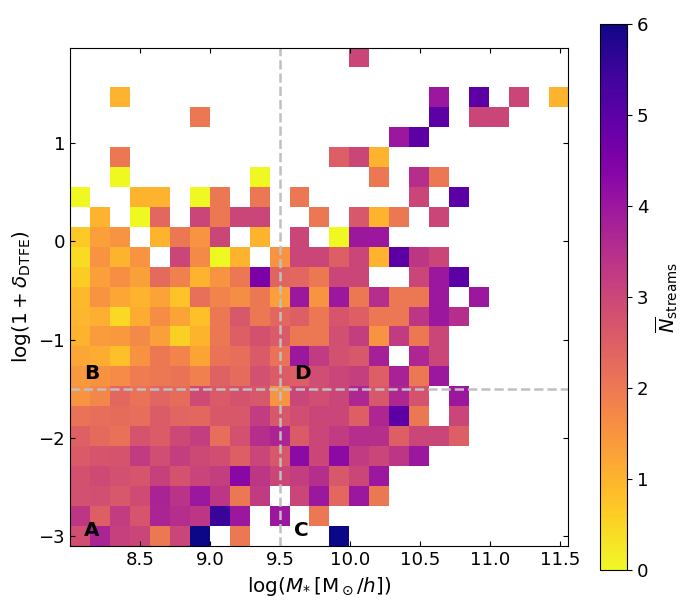}\includegraphics[width=0.5\textwidth]{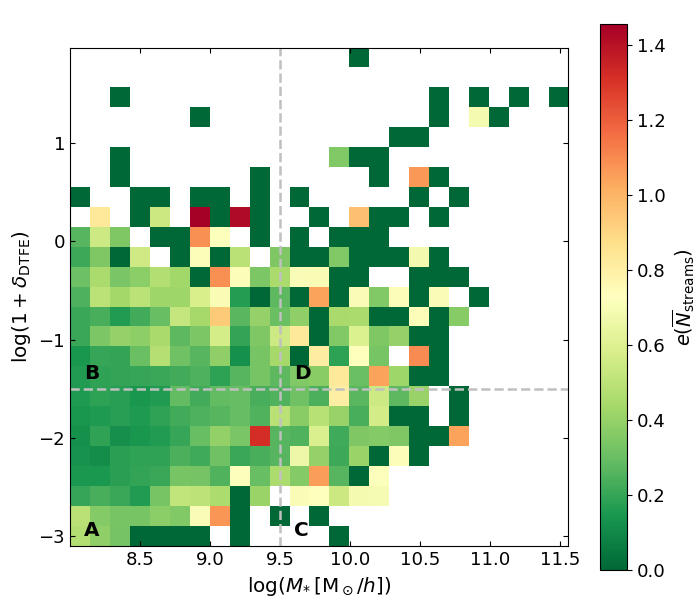}
    \caption{Connectivity variations as a function of mass and local over-density. \textit{Left:} Mean connectivity $\overline{N}_\mathrm{streams}$ in the galaxy mass vs local over-density plane. The pixel colours represent the mean number of streams in a given mass and \LOD~bin. The dashed grey lines show the limits of the four different galaxy populations we study in Sect.~\ref{Sect:SFR}. \textit{Right:} Corresponding bootstrap errors on the means. The dark green pixels (error values of zero) need to be interpreted with caution because they represent bins with only one galaxy (see the number counts in Fig.~\ref{Fig:APP_number2dplane_all}).}
    \label{Fig:rel_Mstar_OV_Nstr_MEANS}
\end{figure*}

Because mass and local density are intrinsically correlated \citep[e.g.][]{AragonCalvo2010}, it is crucial to analyse these two parameters together in order to simultaneously capture their influence on galaxy connectivity. This is done in Fig.~\ref{Fig:rel_Mstar_OV_Nstr_MEANS}, where we present the variation in mean connectivity in the mass-overdensity parameter space (left panel) and the corresponding bootstrap errors (right panel). For reference, the number of galaxies contributing to each pixel of this 2D plane is shown in Fig.~\ref{Fig:APP_number2dplane_all} of Appendix~\ref{Appendix:Number_of_galaxies}.

In addition to the already described trends with galaxy mass, Fig.~\ref{Fig:rel_Mstar_OV_Nstr_MEANS} shows interesting trends with $1 + \delta_\mathrm{DTFE}$. 
For low-mass galaxies (with stellar masses lower than $\sim 10^{9.5}$ $\mathrm{M}_\odot/h$), the mean number of streams strongly decreases with increasing local over-density. The least connected galaxies are located in the highest-density environments (see the yellow region in the top left corner of the plot). Galaxies in these crowded environments are subject to stronger (local) tidal effects \citep[e.g.][]{Hahn2009_tides}, which increase the probability of strong interactions (e.g. by mergers) with respect to galaxies in lower density environments. \cite{AragonCalvo2019_CWdisconnection} has shown that these interactions can lead to the disconnection of galaxies from their filamentary web, thus leading to very low connectivity values. In line with these interpretations, this figure also shows that low-mass galaxies embedded in less crowded regions ($\log(1 + \delta_\mathrm{DTFE}) < -1.5$) have more connections to small-scale filamentary streams. 

In stark contrast with low-mass galaxies, high-mass galaxies ($M_* > 10^{9.5}$ $\mathrm{M}_\odot/h$) do not show any significant trend with local over-density. Their mean connectivity varies between two and five (with few exceptions) regardless of the specific values of mass and density. We note that the tail at the highest $M_*$ and \LOD~values (top right corner) is due to the intrinsic correlation between mass and local environment. High-mass galaxies are less sensitive to the tides driven by the local density, therefore their high connectivity is most probably explained by the trends with mass discussed in Sect.~\ref{SubSect:TrendsMass}.

The right panel of Fig.~\ref{Fig:rel_Mstar_OV_Nstr_MEANS} demonstrates that the results presented in this section are significant because the errors of the relevant pixels are tiny and not correlated with their position in the mass-overdensity plane. Finally, we verified the robustness of 
these results by repeating the same analysis using mass-weighted Delaunay densities (not shown). We found exactly the same trends of connectivity with local density as in Fig.~\ref{Fig:rel_Mstar_OV_Nstr_MEANS}.\\

The local density gives a first-order description of the environment of a galaxy, but it does not encode information on the location of this galaxy in the large-scale environment, set by the different structures of the cosmic web. Knowing the position of a galaxy in the large-scale structures is crucial for fully understanding the results presented in this work. This is shown in the next section.

\section{\label{Sect:envs_LSS}Connectivity in different cosmic web environments}

In this section, we explore the effect of large-scale cosmic environment on galaxy connectivity. It is important to extend the study of environment beyond the first-order analysis of local densities because of the well-established influence of large-scale cosmic tides on matter assembly \citep{Hahn2009_tides, Musso2018_tidesLSS, Paranjape2018_tidesLSS}. Before presenting our results, we recall that information about the local over-density of a galaxy does not allow us to unambiguously determine the position of this object in the cosmic web. This is due to the degeneracies between local and global (cosmic) environments \citep[e.g.][]{Cautun2014}. Figure 13 of Cautun et al. clearly illustrates this point, as the $1+ \delta$ distributions of matter in the cosmic environments of nodes, filaments, walls, and voids largely overlap.\\

\begin{table}[]
    \centering
    \begin{tabular}{c | c | c c c c}
      & Total & A & B & C & D \\ \hline \hline
    All cosmic environments & 2942 & 1750 & 834 & 149 & 209  \\ \hline
    Voids + Walls & 1211 & 981 & 172 & 44 & 14\\
    Filament outskirts & 454 & 281 & 138 & 15 & 20\\
    Filaments & 1213 & 488 & 498 & 86 & 141  \\
    Cluster outskirts & 28 & 0 & 26 & 0 & 2 \\
    Galaxy Clusters & 36 & 0 & 0 & 4 & 32 \\
    \end{tabular}
    \caption{Numbers of galaxies in the different cosmic environments and zones of the mass-overdensity plane (from A to D, see Fig.~\ref{Fig:rel_Mstar_OV_Nstr_MEANS}).}
    \label{Table:number_gal}
\end{table}

\begin{figure}
    \centering
    \includegraphics[width=0.45\textwidth]{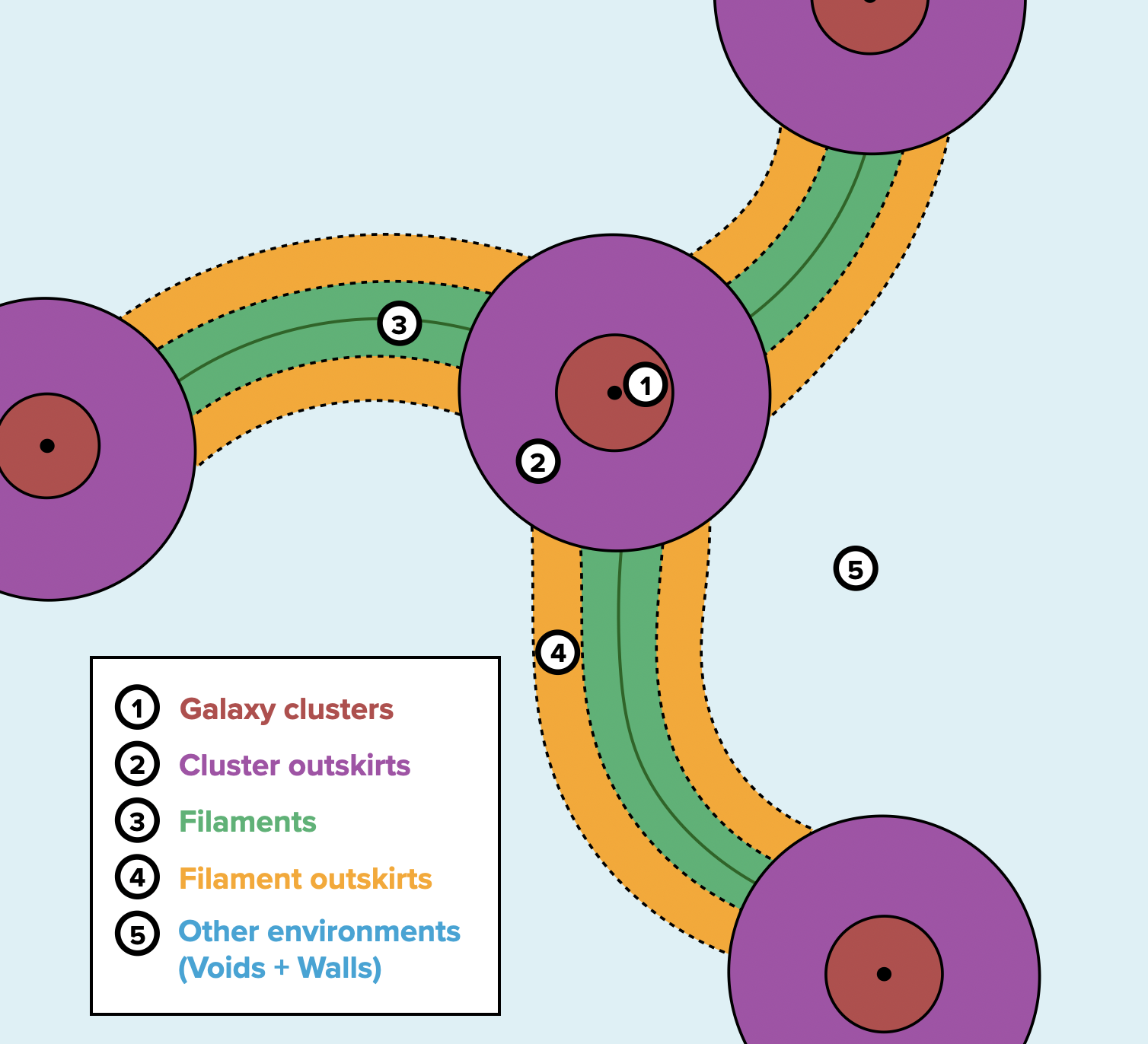}
    \caption{2D illustration of the five different cosmic environments. These are clusters of galaxies (red), cluster outskirts (purple), cosmic filaments (green), filament outskirts (orange), and `other' environments (blue). The exact definitions and corresponding number of galaxies belonging to each environment are presented in the main text.}
    \label{Fig:illustration}
\end{figure}

\begin{figure*}
    \centering
    \includegraphics[width=1\textwidth]{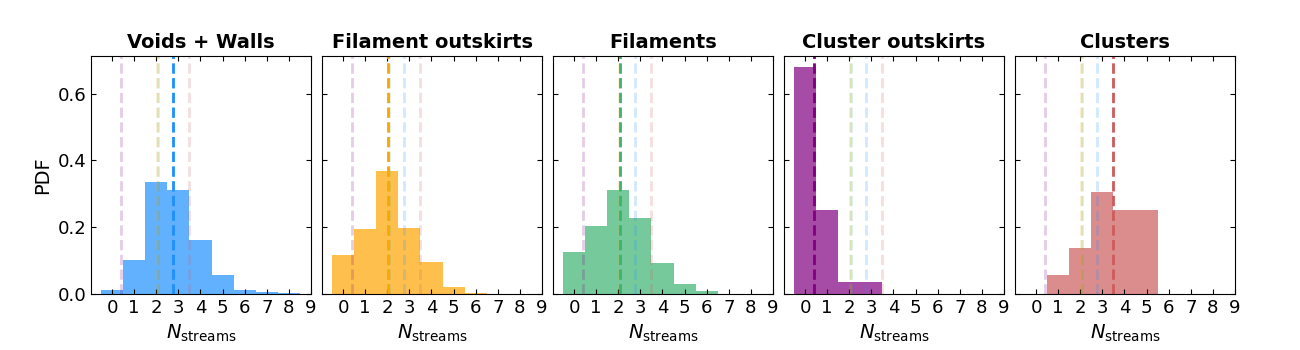}
    \caption{Connectivity distribution as a function of cosmic environments. The five cosmic environments are defined in the main text. The vertical lines represent the mean values in each of the different environments. These are $\overline{N}_\mathrm{streams} =$ 
    $2.77 \pm 0.03, 2.04 \pm 0.06, 2.09 \pm 0.04, 0.43 \pm 0.14,$ and $3.50 \pm 0.19$  for voids and walls, filament outskirts, filaments, cluster outskirts, and clusters, respectively.}
    \label{Fig:Nstr_dist_CW5}
\end{figure*}

\begin{figure*}
    \centering
    \includegraphics[width=1\textwidth]{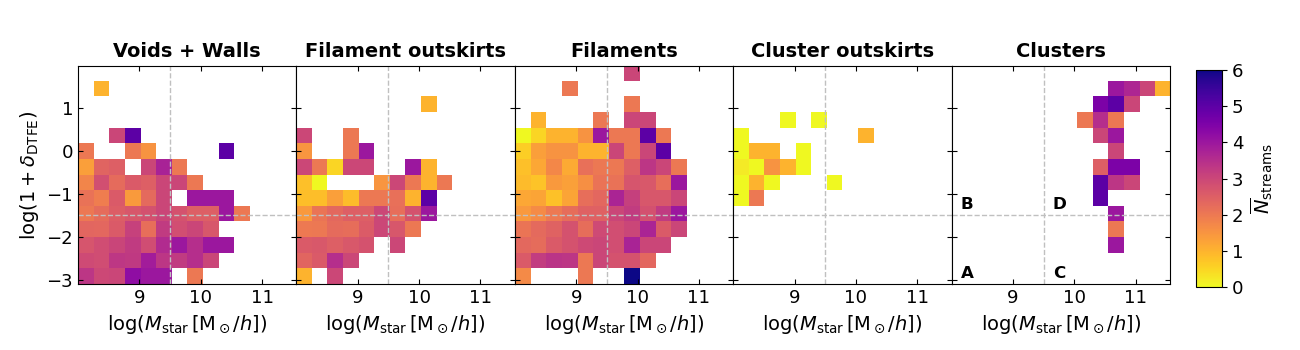}
    \includegraphics[width=1\textwidth]{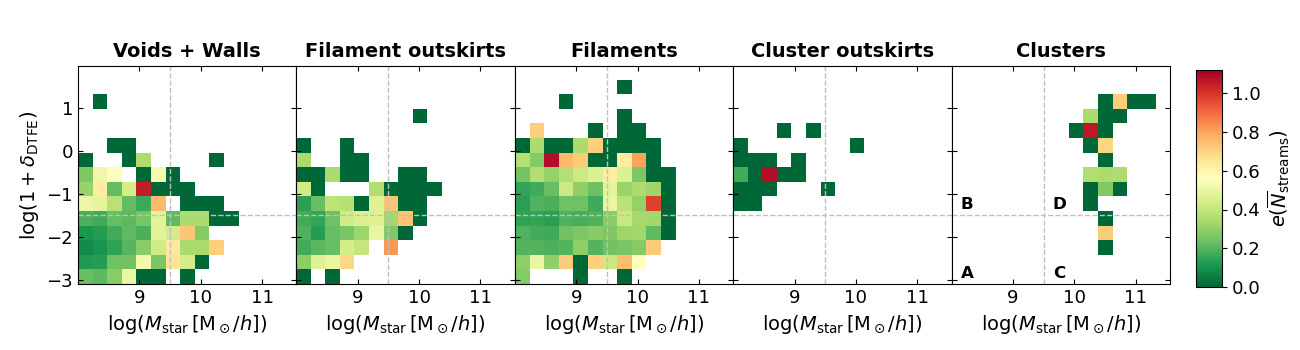}
    \caption{\textit{Top:} Mean galaxy connectivity (shown by the pixel colours) in the mass vs $1 + \delta_\mathrm{DTFE}$ plane as a function of cosmic environments (from left to right panels). The dashed grey lines show the limits of the four different galaxy populations studied in Sect.~\ref{Sect:SFR}. \textit{Bottom:} Corresponding bootstrap errors. The dark green pixels (error values of zero) need to be interpreted with caution as they represent bins with only one galaxy (see the number counts in Fig.~\ref{Fig:APP_number2dplane}).}
    \label{Fig:rel_Mstar_OV_Ntr_CW}
\end{figure*}

We associate galaxies with one of the five different cosmic environments presented in the illustration of Fig.~ \ref{Fig:illustration}. 
The five cosmic environments are defined below, and the number of galaxies in each is reported in the first column of Table~\ref{Table:number_gal}.

First, galaxy clusters are spheres with a radius $R_{200}$ centred on the positions of the FoF haloes with masses $M_{200}>10^{12} \, \mathrm{M}_\odot/h$. Second, cluster outskirts are defined as spherical shells with an inner and outer radius 1 and $3 \times R_{200}$, centred on the positions of galaxy clusters. Third,
cosmic filaments are cylinders aligned with the spine of the (large-scale) skeleton detected in Sect.~\ref{SubSect:Extracting_CW}, and have a radius of 1 cMpc/h. This value was chosen in order to select the regions associated with the cores of cosmic filaments \citep{GalarragaEspinosa2022}.
Filament outskirts are the regions between 1 and 2 cMpc/h from the axis of cosmic filaments (without filament cores). Finally, void and wall environments are all the other regions that do not belong to one of the four described above. We note that we here analyse galaxies in voids and walls together because only little information is available about the physical properties of these cosmic structures (e.g. wall average thickness or void size). This information is required in order to associate galaxies with the structures of the cosmic web described by DisPerSE.\\

Figure~\ref{Fig:Nstr_dist_CW5} presents the connectivity distribution of galaxies split according to these five cosmic environments. The distributions are clearly different, demonstrating a dependence of connectivity on the location of the galaxy in the cosmic web. The corresponding trends in the mass versus \LOD~plane are exhibited in Fig.~\ref{Fig:rel_Mstar_OV_Ntr_CW}, where the mean connectivity and errors are presented in the top and bottom panels, respectively. For completeness, the number of galaxies in each bin of this 2D parameter space is shown in the 2D histogram of Fig.~\ref{Fig:APP_number2dplane}.
We observe the following trends with cosmic environment.

First, in cosmic filaments and filament outskirts, low-mass galaxies in high-density regions (zone B, top left corner) are significantly less connected than the same galaxies in voids and walls. In filaments, voids, and walls, the mean connectivity of these galaxies is $\overline{N}_\mathrm{streams} = 1.43 \pm 0.05$ and $2.34 \pm 0.09$, respectively, yielding a $8.48\sigma$ difference between these cosmic environments. This result can be explained by the different strengths of the cosmic tidal flow \citep{Kraljic2019}. Due to the stronger gravitational pull, galaxies in filaments and their outskirts are subject to stronger large-scale tides than their analogues in walls and voids. For example, \cite{Jhee2022_filaments} presented a clear illustration of halo-mass tidal stripping by dense cosmic filaments. As argued by \cite{Hahn2009_tides} and already mentioned in Sect.~\ref{SubSect:LOD}, strong tides (whether local or cosmic) can prevent the convergence of matter flows onto galaxies and hence the formation of coherent streams. Interestingly, because the galaxies in zone B share local over-density values, the observed decrease in connectivity in large-scale filaments is most probably the result of cosmic tides combined with strong interactions with the environment, which can strip these low-mass galaxies from their streams \citep{AragonCalvo2019_CWdisconnection}.

The interpretation of the very low connectivity values observed in cluster outskirts is much less straightforward because the statistics in these regions is poor. We nevertheless comment on the fact that cluster outskirts are unique environments at the intersection between cosmic filaments and clusters, so that galaxies with different histories co-exist in these regions \citep[e.g. galaxies falling through filaments, splash-back galaxies, or galaxies in groups, as studied in][]{Kuchner2022_inventoryGalaxies, Borrow2023_splashback, Hough2023_clusteroutskirts}. In addition, the question of how galaxies are accreted into cluster cores and the physical processes they undergo during their infall is currently under active investigation \citep[e.g.][and references therein]{Gouin2022, Kotecha2022_fils_clusters, Salerno2022_clusteroutskirts}. At this stage we can therefore only argue that results in cluster outskirts might be a combined effect of galaxy diversity and interactions in this unique environment, but a study with a larger number of galaxies is required.

In stark contrast with the previously studied cosmic environments, Figs.~\ref{Fig:Nstr_dist_CW5} and \ref{Fig:rel_Mstar_OV_Ntr_CW} show that galaxy clusters host systems with the highest connectivity values of all, with a total average of 3.5 streams. Because these cosmic structures dominate the local gravitational field, they are rather insensitive to the cosmic tidal flows. The great number of streams of galaxies in clusters is therefore driven by the high galaxy masses found in these cosmic structures, following the trends presented in Sect.~\ref{SubSect:TrendsMass}.\\

The results of this section echo the analysis in the zoom-in simulations of \cite{Borzyszkowski2017_ZOMG1, RomanoDiaz2017_ZOMG2} and \cite{Garaldi2018_ZOMG3}. In these papers, the authors focused on a few selected haloes, and separated the accreting from stalled ones, finding that their different mass-assembly histories are explained by the location of the halo in the cosmic web \citep[see e.g. Fig. 10 of][]{Borzyszkowski2017_ZOMG1}. While a careful study of accretion rates and outflows along the galactic streams will be done in a follow-up project, from Figs.~\ref{Fig:Nstr_dist_CW5} and \ref{Fig:rel_Mstar_OV_Ntr_CW} one can already hint that accreting haloes might be highly connected objects residing in cosmic environments in which the tidal field is relatively weak, whereas the stalled haloes might rather be disconnected from their matter supply and be embedded in structures where the cosmic flow is strong (e.g. in large-scale filaments).

\section{\label{Sect:SFR}Impact on star formation}

After studying the connectivity of galaxies and understanding its dependencies on mass and environment, we present in this section a first exploration of the impact of galaxy connectivity on star formation. This is a crucial analysis because the material for star formation (cold and dense gas) is predicted to be accreted onto the galaxy via the small-scale streams \citep[e.g.][]{Keres2005, Ocvirk2008, Dekel2009_coldstreams}, such as we detected and studied here. While a more comprehensive analysis including studies of mass-accretion rates and gas properties of the filamentary streams is left for a follow-up project, we can already try to identify any possible effects solely driven by topology here, that is, by the number of connections of the galaxy to filamentary streams. 

\begin{figure}
    \centering
    \includegraphics[width=0.5\textwidth]{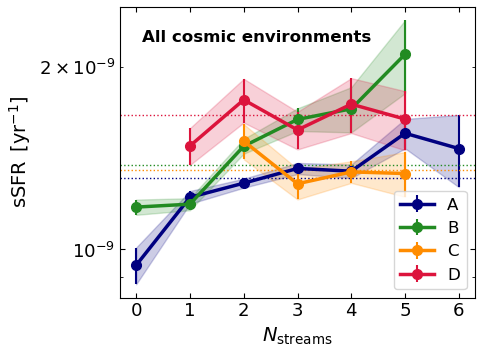}
    \caption{Influence of galaxy connectivity on star formation. The curves show the mean galaxy sSFR as a function of connectivity for the four different galaxy populations from A to D presented in Fig.~\ref{Fig:rel_Mstar_OV_Nstr_MEANS}. The horizontal lines show the average sSFR of all the galaxies in a given population, regardless of the connectivity value. We note that the \textit{y}-axis is in logarithmic scale.}
    \label{Fig:sSFR_ALLCOSMIC}
\end{figure}

In order to break the well-known degeneracies between star formation, galaxy mass, and local density and to probe the specific effects of connectivity, we separated galaxies into the four different populations presented in the mass-overdensity plane of Fig.~\ref{Fig:rel_Mstar_OV_Nstr_MEANS} (see the dashed grey lines). From A to D, galaxies increase in local density and mass. The limits between populations are $M_* = 10^{9.5} \, \mathrm{M}_\odot /h$ and $1+\delta_\mathrm{DTFE} = 10^{-1.5}$ , and the number of galaxies in each is reported in the first line of Table~\ref{Table:number_gal}.

Figure \ref{Fig:sSFR_ALLCOSMIC} presents the variation in mean sSFR as a function of the galaxy connectivity for these four galaxy populations in all cosmic environments combined. For reference, the average sSFR of all the galaxies in a given population is marked by the dotted horizontal lines. For low-mass galaxies, the highest sSFR values are associated with the largest number of connections, yielding a clear positive correlation between star formation and connectivity (see populations A and B, shown in blue and green, respectively). The significance of this relation is estimated using Eq.~\ref{Eq:significance} and is found to be as high as $5.84 \sigma$ and $5.92 \sigma$ for populations in A and B, respectively. This strong sSFR enhancement driven by connectivity is in line with the so-called cold accretion mode introduced in \cite{Keres2005}. Namely, the haloes hosting low-mass galaxies may not be massive enough to support shocks, enabling the cold gas flowing along the filamentary streams to reach the centre of the halo, thus feeding the central galaxy with material for star formation. Consequently, the more streams, the higher the sSFR enhancement. 

On the other hand, for high-mass galaxies (populations C and D, shown in yellow and red, respectively), the relation between the sSFR and connectivity is rather flat. This indicates that star formation in massive galaxies is less dependent on the number of connections of the galaxy to the matter reservoirs outside the halo. In line with the so-called hot accretion mode \citep{Keres2005}, this indicates that in massive systems, star formation might have little to do with potential inflows of cold gas via the filamentary streams, and might instead be regulated by internal processes, such as the recycling of gas within the halo, or the cooling of gas that has been shock-heated by accretion into the halo.
In this scenario, a more important parameter to understand star formation in massive galaxies could be the cooling rate of gas, rather than the galaxy connectivity.\\

\begin{figure}
    \centering
    \includegraphics[width=0.5\textwidth]{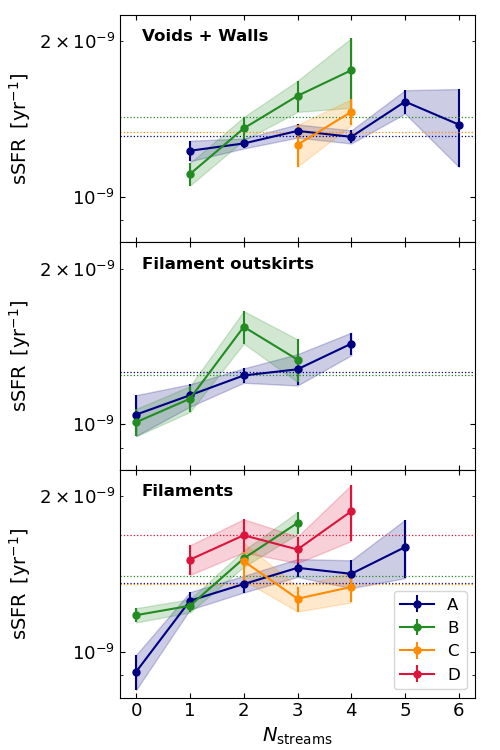}
    \caption{Influence of galaxy connectivity on star formation. The curves show the mean galaxy sSFR as a function of connectivity for different galaxy populations (from A to D, see Fig.~\ref{Fig:rel_Mstar_OV_Nstr_MEANS}) and cosmic environments. The horizontal lines show the average sSFR of a given galaxy population and cosmic environment, regardless of the connectivity value. We note that the \textit{y}-axis is in logarithmic scale.}
    \label{Fig:sSFR_ALL}
\end{figure}

It is established that galaxy properties can be impacted by the large-scale environment \citep[e.g.][and references therein]{Hahn2007a_prop_haloes_LSS, Hahn2007b_EVOprop_haloes_LSS, Laigle2015_vortices, Borzyszkowski2017_ZOMG1, Musso2018_tidesLSS, Paranjape2018_tidesLSS, Malavasi2022_spin}, therefore we further differentiated galaxies with respect to their location in the cosmic web. Figure \ref{Fig:sSFR_ALL} captures the specific role of the large-scale environment on the relation between star formation galaxy connectivity for voids and walls, filament outskirts, and filaments. We refrained from performing this study in clusters and cluster outskirts due to the very low number of galaxies in these structures, as exposed in Table~\ref{Table:number_gal}. Moreover, we note that in order to have statistically meaningful results, bins of \Nstream~with fewer than ten galaxies were removed from this plot (they usually correspond to extreme connectivity values). This figure shows the same qualitative results as in Fig.~\ref{Fig:sSFR_ALLCOSMIC}, that is, the sSFR of low-mass galaxies is largely enhanced with connectivity, while that of high-mass galaxies shows a much milder relation with the number of connected streams.

\begin{figure}
    \centering
    \includegraphics[width=0.5\textwidth]{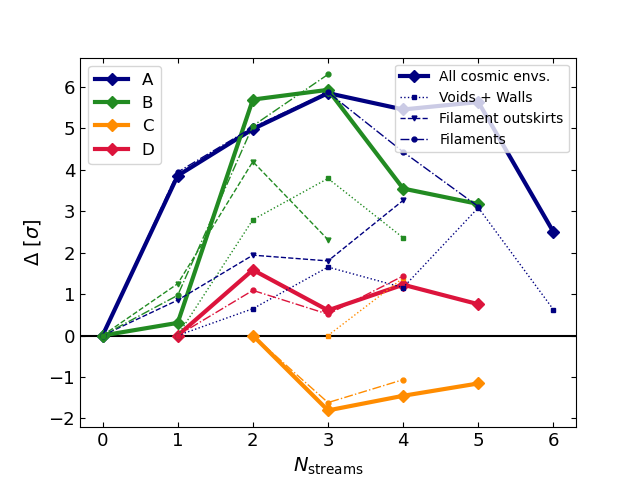}
    \caption{Significance of the sSFR enhancement due to galaxy connectivity. These curves show the results of Eq.~\ref{Eq:significance} for different galaxy populations (see colours) embedded in different cosmic environments (different line styles).}
    \label{Fig:SIGNIFICANCE}
\end{figure}

Nevertheless, the strength of the observed trends strongly varies in the different cosmic structures. This is quantified in Fig.~\ref{Fig:SIGNIFICANCE}, which presents the significance $\Delta$ of the sSFR enhancement due to connectivity. The $\Delta$ values are estimated by
\begin{equation}\label{Eq:significance}
    \Delta(N) = \frac{ \mathrm{sSFR}(N) - \mathrm{sSFR}(N_\mathrm{min}) }{ \sqrt{ \sigma_\mathrm{sSFR}(N)^2 + \sigma_\mathrm{sSFR}(N_\mathrm{min})^2 } },
\end{equation}
where $\mathrm{sSFR}$ and $\sigma_\mathrm{sSFR}$ denote the mean sSFR values and corresponding bootstrap errors as seen in Figs.~\ref{Fig:sSFR_ALLCOSMIC} and \ref{Fig:sSFR_ALL}, respectively, and $N_\mathrm{min}$ represents the lowest number of streams for galaxies in a given population and cosmic environment.

It is striking to see that cosmic filaments (dot-dashed lines with circles) are the places in which the star formation of low-mass galaxies is most enhanced (with up to $6.30\sigma$ for population B). While still significant, this enhancement is more moderate in other cosmic environments, with maximum $\Delta$ values of $3.08\sigma$ in walls and voids, and $4.19\sigma$ in filament outskirts. 
These differences illustrate how the matter reservoirs of the different cosmic environments play an important role in boosting galaxy star formation. At fixed connectivity values, the small-scale streams attached to galaxies embedded in (large-scale) cosmic filaments benefit from the larger matter reservoirs proper to these environments, and are thus probably more efficiently fueled than those in the emptier environments of walls and voids, for instance. 
To summarise, the results in this section show that high connectivity values in matter-rich large-scale environments significantly favour the star formation activity of low-mass galaxies at $z=2$.

\section{\label{Sect:Conclusions}Summary and conclusions}

The question of how galaxies acquire the material from the cosmic web to fuel star formation is fundamental to galaxy evolution. We presented the first comprehensive characterisation of the galaxy connectivity (i.e. the number of filamentary streams attached to a galaxy) in relation with the cosmic environment. We also showed the first steps towards assessing the impact of this topological property on the galaxy SFR. By performing a statistical analysis of 2942 massive ($M_* > 10^{8}$ $\mathrm{M}_\odot / h$) centrals in the TNG50-1 simulation at $z=2$, we reached the main conclusions summarised below.

\begin{itemize}
    \item \textit{(i)} The total connectivity distribution (Fig.~\ref{Fig:Nstream0}) spans a broad range from zero to nine streams. Most of the galaxies ($> 50\%$) are connected to two or three streams, and fewer than $5\%$ of them are connected to five streams or more.\\
    
    \item \textit{(ii)} Galaxy connectivity strongly depends on galaxy mass. We found that low-mass galaxies are less connected than high-mass galaxies on average (Fig.~\ref{Fig:Nstr_dist_5binsM}). Empirically, we established the following simple relation between mean connectivity and galaxy mass: $\overline{N}_\mathrm{streams} \propto 0.5 \,\, \log(M_* \, [\mathrm{M}_\odot/h])$, presented in Fig.~\ref{Fig:meanNstr_binsM}.\\
    
    \item \textit{(iii)} Galaxy connectivity also depends on local environment, with differences between low- and high-mass galaxies (Fig.~\ref{Fig:rel_Mstar_OV_Nstr_MEANS}). We found that low-mass galaxies (with stellar masses lower than $\sim 10^{9.5}$ $\mathrm{M}_\odot/h$) in high local over-density environments are connected to significantly smaller numbers of streams than galaxies of the same mass that are located in lower over-dense regions. This trend with local environment was interpreted by the influence of the stronger tidal forces felt by low-mass galaxies in high over-density environments \citep{Hahn2009_tides, AragonCalvo2019_CWdisconnection}. We showed for high-mass galaxies that their connectivity is independent of local over-density, and that their greater number of connected streams is probably driven by their mass.\\
    
    \item \textit{(iv)} By further disentangling galaxies in different cosmic environments, we found that the average galaxy connectivity decreases from cosmic voids and walls to filament outskirts, from the latter to filament cores, and is the lowest among all in cluster outskirts  (Fig.~\ref{Fig:rel_Mstar_OV_Ntr_CW} and \ref{Fig:Nstr_dist_CW5}). This decrease might be due to the increasing strength of cosmic tides in these cosmic environments \citep[e.g.][]{Musso2018_tidesLSS, Paranjape2018_tidesLSS,  Kraljic2019}. On the other hand, we showed that the average galaxy connectivity is highest of all in galaxy clusters, where the most massive galaxies reside.\\
    
    \item \textit{(v)} We found that galaxy connectivity significantly enhances (up to $\sim 6\sigma$) the star formation of low-mass galaxies, but no significant effect is seen in high-mass galaxies (Figs.~\ref{Fig:sSFR_ALLCOSMIC}). This indicates different dominant accretion modes in low- and high-mass galaxies. \\

    \item \textit{(vi)} We showed that if they keep the connections despite the strong tides, low-mass galaxies in matter-rich regions of the cosmic web (e.g. cosmic filaments) present stronger star formation activities than their analogues in emptier large-scale environments (Fig.~\ref{Fig:SIGNIFICANCE}). This explicitly shows the importance of the large-scale matter reservoirs in fueling the star formation of low-mass galaxies.    
\end{itemize}

These results draw a picture in which star formation is linked to an external parameter describing topology, the galaxy connectivity. 
Within this picture, many connected streams might favour the accretion of cold material from the large scales and thus boost the galaxy star formation, especially in the case of low-mass galaxies. As mentioned in the main body of the paper, it remains to be investigated whether galaxy connectivity is a fundamental parameter or rather a proxy for gas accretion rates, for instance. For example, it remains to be determined whether all the DM streams actively transport matter towards the galaxy, what fraction of gas accreted via the streams is with respect to an isotropic accretion, and more fundamentally, whether mass is the result of connectivity (because of an efficient accretion of matter through the streams) or if the connectivity is driven by mass.
These questions will be answered in the next parts of this series of papers, where we will also investigate the gas properties of the streams.\\

Moreover, throughout this paper, we showed that cosmic filaments host galaxies with the most diverse ranges of masses, local densities, and connectivity values (see e.g. the middle panel of Fig.~\ref{Fig:rel_Mstar_OV_Ntr_CW}). Different galaxy populations therefore co-exist in these cosmic environments, which are also are less extreme than those of clusters of galaxies, and present a rich diversity in terms of gas density and temperature \cite[e.g.][]{GalarragaEspinosa2021, GalarragaEspinosa2022}. This diversity makes cosmic filaments an interesting environment for galaxies, in which the evolution of different populations of galaxies in the broader cosmological picture can be studied.


\begin{acknowledgements}
The authors thank the referee for their very constructive comments and suggestions. DGE would like to thank Ra\'ul Angulo, Rüdiger Pakmor, and Nir Mandelker for useful and insightful discussions. We thank Adam L.~Schaefer and C\'eline Gouin for providing comments on the final version of the draft. We also thank the IllustrisTNG team for making their data publicly available, and for creating a user-friendly and complete website.

\end{acknowledgements}

\bibliography{main} 


\begin{appendix}

\section{\label{Appendix:mainsequence}Galaxies in the $M_* - \mathrm{SFR}$ plane}

\begin{figure}[h!]
    \centering
    \includegraphics[width=0.5\textwidth]{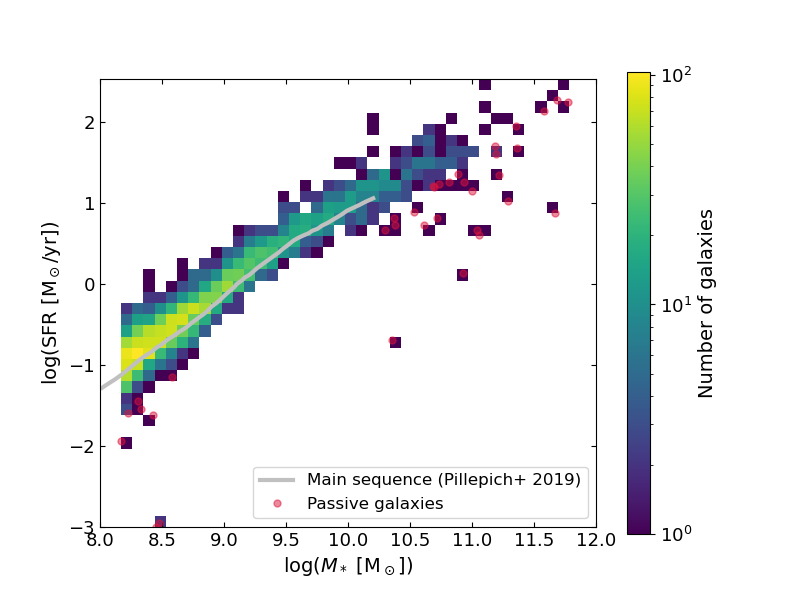}
    \caption{Galaxies in the $M_* - \mathrm{SFR}$ plane.}
    \label{Fig:APP_mainsequence}
\end{figure}

The relation between stellar mass and SFR of the TNG50-1 central galaxies studied in this work is presented in the 2D histogram of Fig.~\ref{Fig:APP_mainsequence}. The silver line shows the main sequence, extracted from \cite{Pillepich2019_TNG50}. We specify that this curve was derived from the study of all the galaxies of the simulation at $z=2$ (centrals and satellites of all masses). Star-forming and passive populations are identified following the method presented in \cite{Pillepich2019_TNG50} (relying on the logarithmic distance to the main sequence). Almost all the galaxies we studied are star forming. Only 35 galaxies of our catalogue are identified as passive (red points in Fig.~\ref{Fig:APP_mainsequence}), which means that the fraction of quenched central galaxies of mass $M_* > 10^{8}$ $\mathrm{M}_\odot / h$ is negligible ($1.2\%$) in TNG50-1 at $z=2$.

Due to the lack of statistics, passive galaxies are not considered in this work (see Sect.~\ref{SubSect:GalaxySelection}). For reference only, we note that the connectivity distribution of these galaxies ranges from zero to five streams, with mean and median values of 2.2 and 2.0, respectively. Roughly half of them lie in clusters (17), 11 are in filaments, and the remaining galaxies are located in the outskirts of filaments and clusters.

\FloatBarrier

\section{\label{Appendix:LargerBox}Connectivity in larger sub-boxes}

In this appendix we show that the size of the sub-boxes we used to detect the small-scale filaments around the central galaxies does not affect the results we presented.
For a random sample of 388 galaxies, we applied the same method as presented in Sect.~\ref{SubSect:Extracting_streams} to DM sub-boxes
with a side of $L = 4$ cMpc/h centred on the galaxy positions. This new value of the box side is one megaparsec larger than the fiducial one and is the largest possible value while maintaining the pixel size (i.e. the resolution) fixed to the original value. The numerical load of larger boxes exceeds the capacity of the DisPerSE code.

\begin{figure}[h!]
    \centering
    \includegraphics[width=0.5\textwidth]{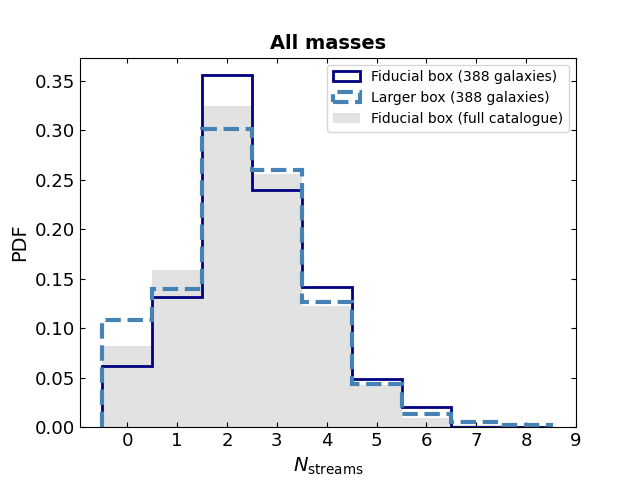}
    \includegraphics[width=0.5\textwidth]{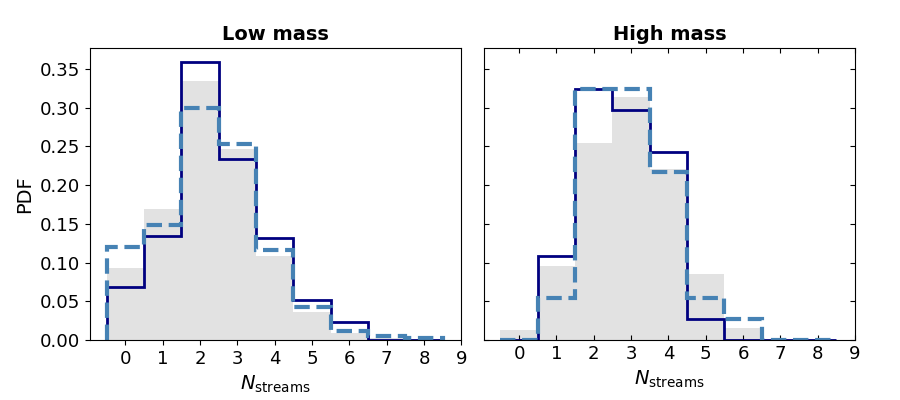}
    \caption{Comparison of connectivity distributions derived from different galaxy box sizes. The fiducial box side is $L = 3$ cMpc/h (grey and solid blue histograms). The larger boxes, in dashed blue, have a side of $L = 4$ cMpc/h. \textit{Top:} Results for galaxies of all masses. \textit{Bottom:} Distributions for galaxies separated into two mass bins. }
    \label{Fig:APP_Nstr_dist_LARGERBOX}
\end{figure}

Figure~\ref{Fig:APP_Nstr_dist_LARGERBOX} compares the resulting connectivity distribution to that derived using the fiducial box size for galaxies of all masses (top panel) and in the low- and high-mass bins (bottom). Following the main text, these bins are defined by the mass limit of $10^{9.5} \, \mathrm{M}_\odot /h,$ and the 388 randomly selected galaxies are split into 351 and 37 low- and high-mass objects, respectively.

The connectivity distributions are essentially the same. This is confirmed by the \textit{p}-value of 0.97 obtained from the two-sample Kolomogorov-Smirnov test comparing the distribution derived from the larger sub-boxes (dashed blue) to the fiducial one (grey) for galaxies of all masses.

\FloatBarrier

\section{\label{Appendix:Disperse_cut}Setting the value of the DisPerSE persistence threshold to extract the small-scale filamentary streams}

\begin{figure*}[h!]
    \centering
    \includegraphics[width=1\textwidth]{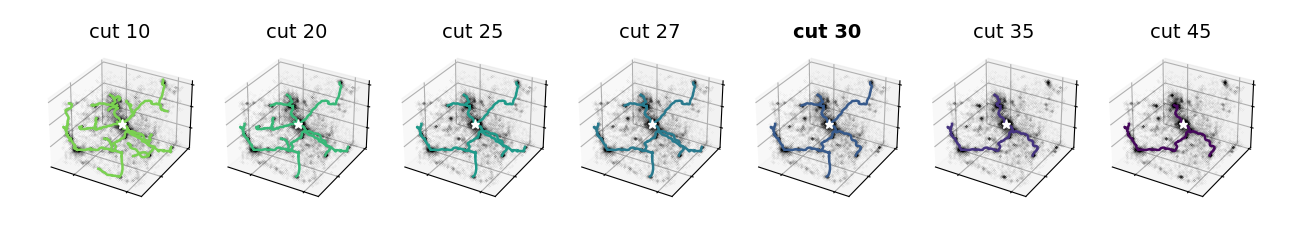}
    \includegraphics[width=1\textwidth]{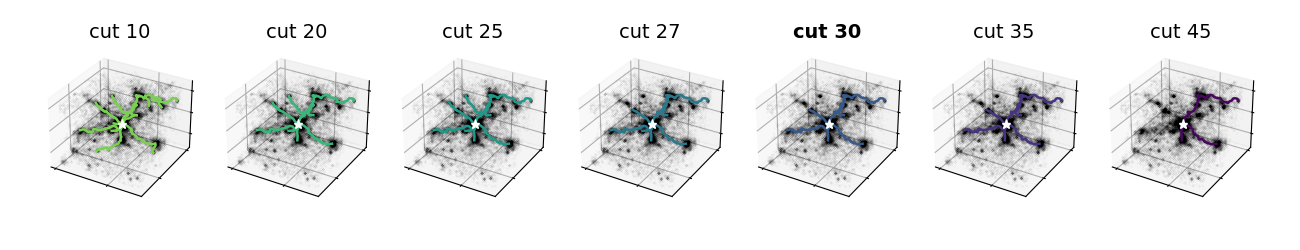}
    \includegraphics[width=1\textwidth]{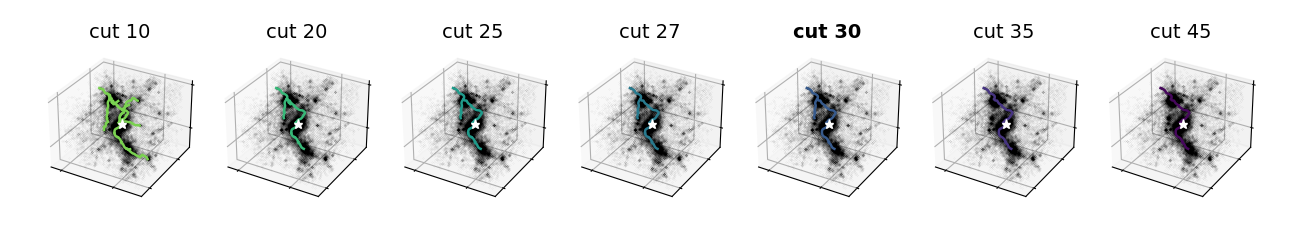}
    \caption{Effect of the DisPerSE persistence threshold (set by the value of the \textit{cut} parameter) on the resulting small-scale skeleton around three random galaxies (from top to bottom). The black points in the background present a random sub-sample 1/1000) of the underlying DM distribution, and the central white star indicates the position of the galaxy. These boxes have a side length of 3 cMpc/h.}
    \label{Fig:APP_3d_skels}
\end{figure*}

\begin{figure}[h!]
    \centering
    \includegraphics[width=0.5\textwidth]{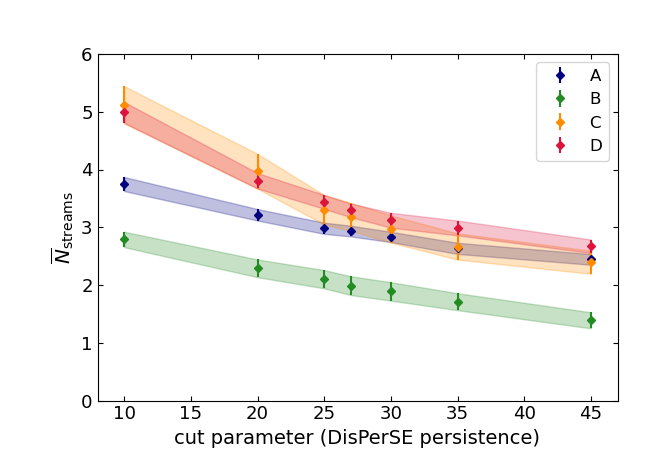}
    \caption{Mean connectivity as a function of persistence for the four galaxy populations defined in the mass-overdensity plane of Fig.~\ref{Fig:rel_Mstar_OV_Nstr_MEANS}.}
    \label{Fig:APP_Nstr_vs_Persistence}
\end{figure}

Figure~\ref{Fig:APP_3d_skels} shows three examples of galaxies for which seven different skeletons were extracted, each with a value of the DisPerSE persistence threshold in the range of \textit{cut} = [10, 45]. As expected, these visual illustrations qualitatively show that the total number of filaments decreases with increasing values of persistence because of the progressive trimming of short branches. Nevertheless, galaxy connectivity is only very mildly affected as these short branches are only rarely connected to the central galaxy.

\begin{figure*}[h!]
    \centering
    \includegraphics[width=1\textwidth]{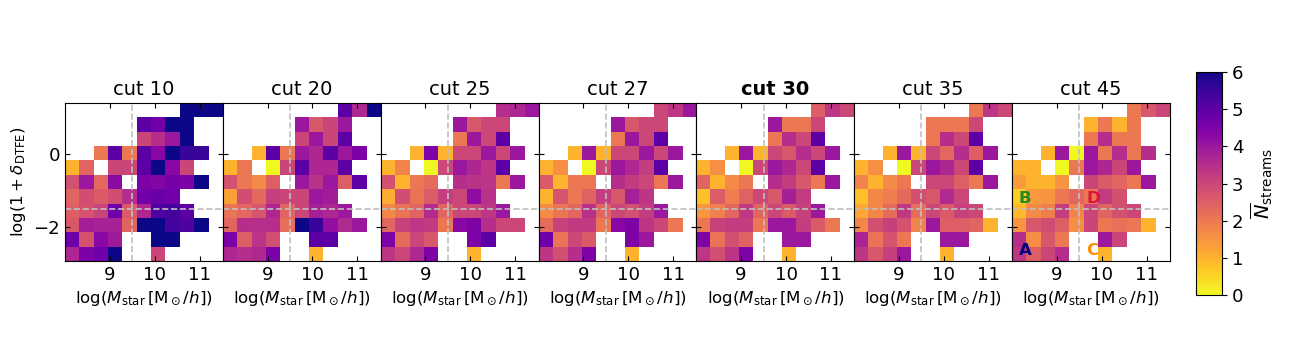}
    \caption{Effect of the DisPerSE persistence threshold on the mean connectivity as a function of galaxy mass and local density. This figure is similar to Fig.~\ref{Fig:rel_Mstar_OV_Nstr_MEANS}, but shows the results for 372 random galaxies.}
    \label{Fig:APP_rel2dplane}
\end{figure*}

In Fig.~\ref{Fig:APP_Nstr_vs_Persistence} we report the evolution of the mean connectivity as a function of persistence for the galaxy populations already introduced in Fig.~\ref{Fig:rel_Mstar_OV_Nstr_MEANS}. The error bars in this figure correspond to the errors on the mean, computed by bootstrap resampling. We note that given the high numerical cost of detecting the streams with DisPerSE, we limited this analysis to a random set of 372 galaxies. This number represents $\sim 13 \%$ of the total galaxy dataset.
The connectivity shows a very shallow linear decrease with increasing persistence (as a consequence of the progressive removal of low-significance branches), and in our experience, deviations from this linear trend arise when noise-induced spurious filaments contribute significantly. Therefore, examining the deviations from linearity in this figure, we show that persistence thresholds above 25 are adequate choices.
An even more granular view is provided in Fig.~\ref{Fig:APP_rel2dplane}, where we show the dependence of Fig.~\ref{Fig:rel_Mstar_OV_Nstr_MEANS} on the chosen persistence value. All panels except the leftmost two show consistent results, albeit with different normalisation for the reason discussed above.


These results thus show that the statistical analysis we performed very weakly depends on the exact value of the DisPerSE persistence threshold, provided spurious filaments are efficiently removed. Therefore, any value of the \textit{cut} parameter above 25 represents an adequate choice.

\section{\label{Appendix:Disperse_cutLSS}Setting the value of the DisPerSE persistence threshold to extract the cosmic skeleton}

Testing the robustness of the positions of the cosmic filaments recovered by DisPerSE is far from being trivial. A true physical reference cannot be easily determined in the simulations, and extracting the skeleton with another detection technique \citep[e.g.][]{Cautun2013nexus, Tempel2014, Bonnaire2020Trex} and comparing the outputs would rather be a characterisation of the similarities and differences of the different methods \citep[already performed by][]{Libeskind2018}, and not a study of the detection of the actual cosmic filaments.

Therefore, in order to test the robustness of the cosmic skeleton and fix the value of the DisPerSE persistence threshold, we focused on the positions of the ending points of the filaments, the CPmax, which by construction correspond to the topological nodes of the skeleton. We compared the positions of the CPmax to those of the most massive FoF haloes of the TNG50-1 simulation. The latter were divided into the following mass bins, which were chosen in order to encompass the main classes of objects:
\begin{itemize}
    \item $M_{200} \in [10^{11}, 5 \times 10^{11}[  \,\, \mathrm{M}_\odot /\mathrm{h}$ (small haloes),
    \item $M_{200} \in [5 \times 10^{11}, 10^{12}[ \,\, \mathrm{M}_\odot /\mathrm{h}$ (groups),
    \item $M_{200} \geq 10^{12} \mathrm{M}_\odot /\mathrm{h}$ (clusters).
\end{itemize}
Table~\ref{Table:APP_match_CPmax_Clusters} presents the results of this comparison for four skeletons detected with different values of the DisPerSE persistence threshold (from 4 to 7). This range encompasses adapted \textit{cut} values given the pixel values of the input DM density grid.
The numbers of this table correspond to the fractions of FoF haloes hosting a CPmax within their $R_{200}$ (first block) and, conversely, the fractions of CPmax coinciding with the positions of the selected FoF haloes (also within $R_{200}$), in the second block.

A robust skeleton maximises the fraction of CPmax in the most massive haloes (by reducing the number of critical points in less massive objects or other less dense environments) while identifying $100\%$ of the most massive haloes as CPmax. This is the case for the skeleton extracted with a persistence threshold of 6, which was therefore adopted for the classification of galaxies in different cosmic environments.

\begin{table*}[h!]
\caption{Fraction of FoF haloes of the TNG50-1 simulation hosting DisPerSE CPmaxs, and conversely.}
\label{Table:APP_match_CPmax_Clusters}     
\centering  
\begin{tabular}{ c | c  c  c  c }
    & cut 4 & cut 5 & cut 6 & cut 7 \\ \hline \hline
  Small haloes hosting CPmax & $26.7\%$ & $18.4\%$ & $13.3\%$ & $9.34\%$ \\
  Groups hosting CPmax & $91.1\%$ & $89.4\%$ & $88.6\%$ & $78.9\%$ \\
  Clusters hosting CPmax & $100\%$ & $100\%$ & $100\%$ & $98.4\%$ \\

  \hline
  CPmax in small haloes & $52.9\%$ & $46.4\%$ & $39.0\%$ & $33.2\%$ \\
  CPmax in Groups & $24.4\%$ & $30.6\%$ & $35.2\%$ & $37.9\%$ \\
  CPmax in Clusters & $13.7\%$ & $17.4\%$ & $20.3\%$ & $24.2\%$ \\
  Total fraction of matched CPmax & $91.1\%$ & $94.4\%$ & $94.5\%$ & $95.3\%$ \\ 
 \end{tabular}
\end{table*}

\section{\label{Appendix:Number_of_galaxies}Distribution of galaxies in the 2D parameter space of mass versus local density}

\begin{figure*}[h!]
    \centering
    \includegraphics[width=0.6\textwidth]{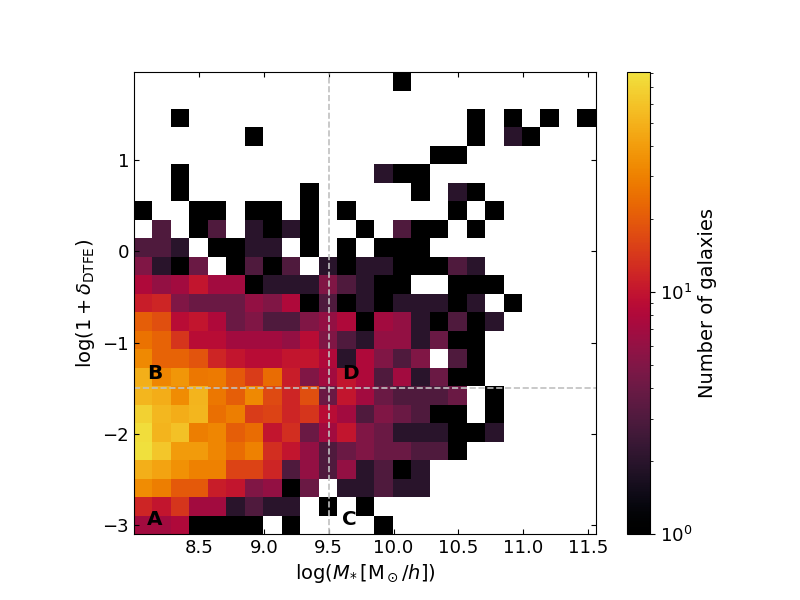}
    \caption{2D histogram of the number of galaxies in the mass-overdensity plane (see Sect.~\ref{SubSect:LOD}).}
    \label{Fig:APP_number2dplane_all}
\end{figure*}

\begin{figure*}[h!]
    \centering
    \includegraphics[width=1\textwidth]{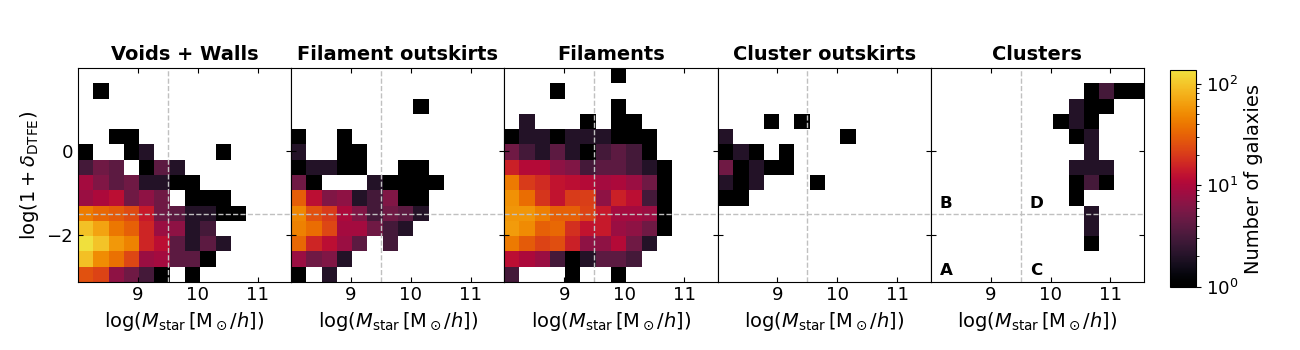}
    \caption{2D histogram of the number of galaxies in the mass-overdensity plane, further segregated by cosmic environment (see Sect.~\ref{Sect:envs_LSS}).}
    \label{Fig:APP_number2dplane}
\end{figure*}

This appendix presents the histograms of the distribution of galaxies in the mass-\LOD~ plane for all the galaxies (Fig.~\ref{Fig:APP_number2dplane_all}), and for galaxies in the different cosmic environments (Fig.~\ref{Fig:APP_number2dplane}).

\end{appendix}

\end{document}